# Perfect imaging with geodesic waveguides

**Juan C. Miñano, Pablo Benítez, and Juan C. González**

Universidad Politécnica de Madrid, Cedint
Campus de Montegancedo
28223 Madrid, Spain

E-mail: jc.minano@upm.es

**Abstract.** Transformation optics is used to prove that a spherical waveguide filled with an isotropic material with radial refractive index $n=1/r$ has radial polarized modes (i.e. the electric field has only radial component) with the same perfect focusing properties as the Maxwell Fish-Eye lens. The approximate version of that device using a thin waveguide with a homogenous core paves the way to experimentally prove perfect imaging in the Maxwell Fish Eye lens.

## 1    Introduction

The Maxwell Fish-Eye (MFE) lens is a positive, isotropic and radial refractive-index distribution which has been well studied in the Geometrical Optics framework. One of the most interesting characteristics of the MFE's refractive index distribution is that any point of the space has a perfect conjugate, *i.e.*, the rays issuing from it are perfectly focused at another point. These two perfect conjugate points are related by an inversion about the origin. Recently, Leonhardt [1] and Leonhardt and Philbin [2] proved that this refractive index distribution has perfect wave-optical properties as well. In particular, Leonhardt [1] proved that two-dimensional Helmholtz scalar-waves generated by a point source are perfectly focused in an "infinitely-well localized drain" located at the conjugate point of the source. The same conclusions were confirmed later using a different approach in [3].

Our purpose here is to show by means of Transformation Optics that we can design a spherical waveguide with the same perfect-focusing properties. We will show that this spherical waveguide is easier to manufacture than the MFE gradient-index distribution, and thus the experimental realization of its properties should be closer. In the final section herein we extend this method to more general two-dimensional radial refractive-index distributions and their corresponding geodesic waveguides.

The paper is organized as follows: Section 1.1 reviews the tools of Transformation Optics that will be employed thereafter, in Section 2, to generate the transformation between the MFE and a spherically symmetric isotropic medium. In Section 3, with the aid of two perfect conductor sheets, the transformation is restricted to the media within two corresponding waveguides. The extension to more general geodesic waveguides is given in Section 4. Finally some conclusions are given in section 5.

### 1.1    Media and coordinate transformations

Maxwell's equations for a current-free, charge-free medium can be written as:

$$\nabla \times \mathbf{E} = -\frac{\partial \mathbf{B}}{\partial t} \qquad \nabla \times \mathbf{H} = \frac{\partial \mathbf{D}}{\partial t} \qquad \nabla \cdot \mathbf{D} = 0 \qquad \nabla \cdot \mathbf{B} = 0 \qquad (1)$$

They are completed with the following constitutive relations wherein $\varepsilon$ and $\mu$ are the 3 x 3 permittivity and permeability tensors:

$$\mathbf{D} = \varepsilon_0 \varepsilon \mathbf{E} \qquad \mathbf{B} = \mu_0 \mu \mathbf{H} \qquad (2)$$

Assume $x_1$, $x_2$, $x_3$ are the usual Cartesian spatial coordinates. A change of the variables $x_1$, $x_2$, $x_3$ can have two different interpretations ([4]-[9]). In the first, which is referred as a media transformation, the new variables are considered to define another Cartesian space $x_1'$, $x_2'$, $x_3'$ (which we will call final space, to distinguish it from the initial space). Then the functions $x_1'(x_1, x_2, x_3)$, $x_2'(x_1, x_2, x_3)$, $x_3'(x_1, x_2, x_3)$ define a mapping between the initial and final spaces. Let's define the vector fields $\mathbf{E}'$, $\mathbf{H}'$, $\mathbf{D}'$ and $\mathbf{B}'$ in this new space by

$$\mathbf{E}' = \Lambda_{xx'}^T \mathbf{E} \qquad \mathbf{H}' = \Lambda_{xx'}^T \mathbf{H}$$
$$\mathbf{D}' = \varepsilon_0 \varepsilon' \mathbf{E}' \qquad \mathbf{B}' = \mu_0 \mu' \mathbf{H}' \qquad (3)$$

where

$$\varepsilon' = \det(\Lambda_{xx'}) \Lambda_{xx'}^{-1} \varepsilon \left( \Lambda_{xx'}^{-1} \right)^T \qquad \mu' = \det(\Lambda_{xx'}) \Lambda_{xx'}^{-1} \mu \left( \Lambda_{xx'}^{-1} \right)^T \qquad (4)$$

and where $\Lambda_{xx'}$ is the Jacobian transformation matrix

$$\Lambda_{xx'} = \begin{pmatrix} \dfrac{\partial x_1}{\partial x_1'} & \dfrac{\partial x_1}{\partial x_2'} & \dfrac{\partial x_1}{\partial x_3'} \\ \dfrac{\partial x_2}{\partial x_1'} & \dfrac{\partial x_2}{\partial x_2'} & \dfrac{\partial x_2}{\partial x_3'} \\ \dfrac{\partial x_3}{\partial x_1'} & \dfrac{\partial x_3}{\partial x_2'} & \dfrac{\partial x_3}{\partial x_3'} \end{pmatrix} \qquad (5)$$

It happens that the vector fields $\mathbf{E}'$, $\mathbf{H}'$, $\mathbf{D}'$ and $\mathbf{B}'$ satisfy the equations

$$\nabla_{x'} \times \mathbf{E}' = -\frac{\partial \mathbf{B}'}{\partial t} \qquad \nabla_{x'} \times \mathbf{H}' = \frac{\partial \mathbf{D}'}{\partial t} \qquad \nabla_{x'} \cdot \mathbf{D}' = 0 \qquad \nabla_{x'} \cdot \mathbf{B}' = 0 \qquad (6)$$

Here the subscripts in the nabla denote taking the derivatives in the Cartesian coordinates $x_1'$, $x_2'$, $x_3'$. From Eq. (6) can be concluded that fields $\mathbf{E}'$, $\mathbf{H}'$, $\mathbf{D}'$ and $\mathbf{B}'$ are electromagnetic fields in the final space with a medium whose permittivity and permeability tensors are $\varepsilon'$ and $\mu'$. In this interpretation we can easily calculate electromagnetic vector fields in one space if we know a electromagnetic vector field in the other space.

The second interpretation of a change of variables $u_1(x_1, x_2, x_3)$, $u_2(x_1, x_2, x_3)$, $u_3(x_1, x_2, x_3)$ is to consider it as describing the same space with the same medium but with a different coordinate system (in general, curvilinear) $u_1$, $u_2$, $u_3$. This can be referred to as a coordinate transformation. The same equations (6), (3) and (5) are formally fulfilled for this coordinate transformation provided that we perform the following substitutions in those equations: the variables $u_1$, $u_2$, $u_3$ substitute for $x_1'$, $x_2'$, $x_3'$; the functions $\mathbf{E}'$, $\mathbf{H}'$, $\mathbf{D}'$ and $\mathbf{B}'$ are substituted for by the "normalized electromagnetic fields" $\hat{\mathbf{E}}$, $\hat{\mathbf{H}}$, $\hat{\mathbf{D}}$ and $\hat{\mathbf{B}}$ (which are actually not true electromagnetic fields), and the tensors $\varepsilon'$ and $\mu'$ by the "normalized tensors" $\hat{\varepsilon}$ and $\hat{\mu}$ (which are not true permittivity and permeability tensors either). The relationship of the normalized field $\hat{\mathbf{E}}$ with the actual electric field $\mathbf{E}$ in the new coordinate system (*i.e.*, $\mathbf{E} = E_1 \mathbf{u}_1 + E_2 \mathbf{u}_2 + E_3 \mathbf{u}_3$; $\mathbf{u}_1$, $\mathbf{u}_2$, $\mathbf{u}_3$ are the 3 unit vectors of this new coordinate system) is $\hat{\mathbf{E}} = \mathsf{H}_u \mathbf{E}$, where $\mathsf{H}_u$ is the diagonal matrix diag($h_1, h_2, h_3$) and $h_i$ are the scale factors $h_i^2 = (\partial x_1/\partial u_i)^2 + (\partial x_2/\partial u_i)^2 + (\partial x_3/\partial u_i)^2$. The same relationship applies for the normalized $\hat{\mathbf{H}}$ and actual $\mathbf{H}$ magnetic fields.

The transformation formulas between two spaces with arbitrary coordinate systems $u_1$, $u_2$, $u_3$ and $u_1'$, $u_2'$, $u_3'$ can be obtained by chaining the media and coordinate transformations. When the two coordinate systems (initial and final) are orthogonal, the resulting formulas are much simpler that in the general case [10]: the permittivity and permeability tensors of the original medium in the first orthogonal system are related with the permittivity and permeability tensors of the final medium, in the second orthogonal system, by

$$\varepsilon' = \frac{1}{\det(M)} M \varepsilon M^T \qquad \mu' = \frac{1}{\det(M)} M \mu M^T \qquad (7)$$

Here $M^T$ denotes the transpose of the matrix M, which is

$$M = \begin{pmatrix} \frac{h'_1}{h_1} \frac{\partial u'_1}{\partial u_1} & \frac{h'_1}{h_1} \frac{\partial u'_1}{\partial u_2} & \frac{h'_1}{h_1} \frac{\partial u'_1}{\partial u_3} \\ \frac{h'_2}{h_2} \frac{\partial u'_2}{\partial u_1} & \frac{h'_2}{h_2} \frac{\partial u'_2}{\partial u_2} & \frac{h'_2}{h_2} \frac{\partial u'_2}{\partial u_3} \\ \frac{h'_3}{h_3} \frac{\partial u'_3}{\partial u_1} & \frac{h'_3}{h_3} \frac{\partial u'_3}{\partial u_2} & \frac{h'_3}{h_3} \frac{\partial u'_3}{\partial u_3} \end{pmatrix} \qquad (8)$$

Here $h_i'$ are the scale factors of the variables $u_i'$. The fields **E** (and **H**) of the original medium in the first orthogonal system are related to **E'** (and **H'**) of the transformed medium in the second orthogonal system by (see Appendix)

$$\mathbf{E} = M^T \mathbf{E'} \qquad \mathbf{H} = M^T \mathbf{H'} \qquad (9)$$

Note that none of the vectors and tensors in Eq. (7) and (9) are normalized, in the sense previously defined.

## 2 Transformation from MFE to a spherical medium

### 2.1 Initial space

The two-dimensional analysis of the MFE by Leonhardt [1] describes the propagation of linear polarized waves in an isotropic cylindrical medium in which the electric field vector **E** is parallel to the cylinder-axis, and the point sources and drains become lines. That analysis, however, also describes propagation of linearly polarized fields in the z-axis for more general media, not necessarily isotropic or cylindrical. For instance, it can be easily checked that the time-harmonic field $\mathbf{E} = E(x,y)e^{-i\omega t}\mathbf{z}$, the modulus of which, $E(x,y)$, fulfils the following 2D-Helmholtz equation (Eq. (6) in [1])

$$\Delta E + k^2 n^2 E = 0 \qquad (10)$$

is a solution of Maxwell's equations (1) for an anisotropic medium with permittivity and permeability tensors of the form $\varepsilon = \text{diag}(\varepsilon_x, \varepsilon_y, \varepsilon_z)$, $\mu = \text{diag}(\mu_x, \mu_y, \mu_z)$, where

$$\varepsilon_z \text{ independent of } z \qquad \mu_x = \mu_y = \mu_\perp = \text{constant} \qquad (11)$$

and $k=\omega/c$ and $n^2=\varepsilon_z \mu_\perp$. Therefore, the waves described by Leonhardt in [1] also propagate in that anisotropic medium, provided that $\varepsilon_z$ and $\mu_\perp$ also fulfill

$$\sqrt{\varepsilon_z \mu_\perp} = n(\rho) = \frac{2}{1+\rho^2} \qquad (12)$$

Here $\rho^2=x^2+y^2$. Note that the elements $\varepsilon_x$, $\varepsilon_y$, and $\mu_z$ may depend on z, showing that this medium is not necessarily cylindrically symmetric.

We are looking for a transformation from a space with a medium of this type (that propagates the 2D-Helmholtz fields of the 2D-MFE) to another final space with a simpler medium (with spherical symmetry). The specific selection of the initial medium will be done later, once the transformation between the two spaces is described.

*2.2 The transformation*

Consider cylindrical coordinates $\rho, \psi, z$ for the initial medium and spherical coordinates $r, \theta, \phi$ for the final medium (we have dropped the primes of the $r, \theta, \phi$ variables to simplify our nomenclature). These three functions define the transformation:

$$\rho = \tan\frac{\theta}{2} \qquad \psi = \phi \qquad z = f(r) \qquad (13)$$

The two first equations of Eq. (13) defines a stereographic projection (see Fig. 1), mapping points of the plane $z$=constant onto points of the sphere $r$=constant. This stereographic projection is not exactly the same as that used in Ref [1], where North was the projection pole. The present stereographic projection simplifies some of the following equations. The transformation equations of the North Pole stereographic projection are the same as Eq. (13) if $\theta$ is substituted by $\pi$-$\theta$.

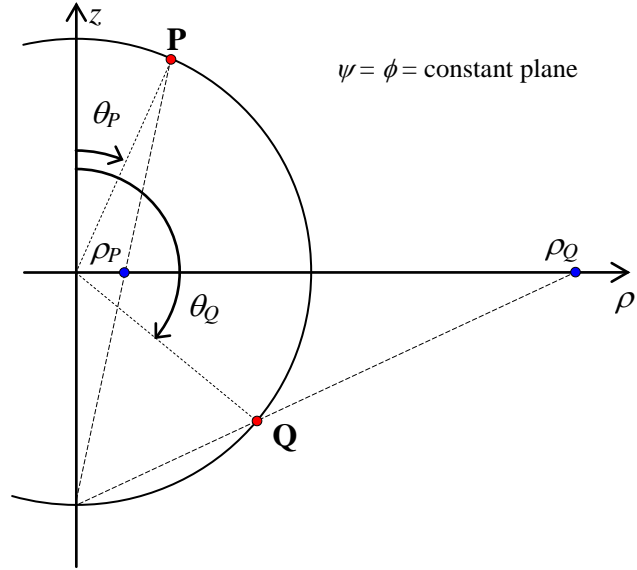

Fig. 1. The stereographic projection maps points of a sphere onto points of a plane. A point on the sphere **P** and its corresponding point on the plane $z$=0 are aligned with the sphere's South Pole. The plane of the drawing has $\psi = \phi$ = constant.

Eq. (7) and (9) are particularly simple when the transformation is such that it can be written as three functions of a single variable, which is the case for Eq. (13). Taking into account that the scale factors of the cylindrical and spherical coordinates are $h_\rho$=1, $h_\psi$=$\rho$, $h_z$=1 and $h_r$=1, $h_\theta$=$r$, $h_\phi$=$r\sin\theta$, the matrix M is

$$\mathbf{M} = \begin{pmatrix} \dfrac{h_r}{h_\rho}\dfrac{\partial r}{\partial \rho} & \dfrac{h_r}{h_\psi}\dfrac{\partial r}{\partial \psi} & \dfrac{h_r}{h_z}\dfrac{\partial r}{\partial z} \\ \dfrac{h_\theta}{h_\rho}\dfrac{\partial \theta}{\partial \rho} & \dfrac{h_\theta}{h_\psi}\dfrac{\partial \theta}{\partial \psi} & \dfrac{h_\theta}{h_z}\dfrac{\partial \theta}{\partial z} \\ \dfrac{h_\phi}{h_\rho}\dfrac{\partial \phi}{\partial \rho} & \dfrac{h_\phi}{h_\psi}\dfrac{\partial \phi}{\partial \psi} & \dfrac{h_\phi}{h_z}\dfrac{\partial \phi}{\partial z} \end{pmatrix} = \begin{pmatrix} 0 & 0 & \dfrac{1}{f'(r)} \\ r(1+\cos\theta) & 0 & 0 \\ 0 & r(1+\cos\theta) & 0 \end{pmatrix} \qquad (14)$$

*2.3 Transformed fields*

Let us consider first the case in which we choose $f(r) = \ln(r)$. This selection makes a bijective mapping from $-\infty < z < \infty$ to $r > 0$, covering the whole initial and final spaces. According to Eq. (9), the electric field in the final medium **E'**, as a function of the electric field in the initial medium of **E** (**E**=$E_\rho\boldsymbol{\rho}$+$E_\psi\boldsymbol{\psi}$+$E_z\mathbf{z}$), is given by

$$\mathbf{E}' = E'_r \mathbf{r} + E'_\theta \mathbf{\theta} + rE'_\phi \mathbf{\phi} = \frac{E_z}{r} + \frac{E_\rho \mathbf{\theta} + E_\psi \mathbf{\phi}}{r(1+\cos\theta)} \tag{15}$$

Since we are interested in the z-polarized waves, *i.e.* **E** has only a z-component ($E_\rho=0$, $E_\psi=0$), we obtain that the transformed field in spherical coordinates **E'** has only its r-component:

$$\mathbf{E}' = \frac{E_z}{r}\mathbf{r} \tag{16}$$

Also, its magnitude is proportional to that of the z-polarized wave in the initial medium.

*2.4 Permittivity and permeability tensors of the final medium*

Consider the initial medium described in section 2.1, *i.e.*, with $\varepsilon=\mathrm{diag}(\varepsilon_\rho, \varepsilon_\psi, \varepsilon_z)$, $\mu=\mathrm{diag}(\mu_\perp, \mu_\perp, \mu_z)$ and with $\varepsilon_z$ and $\mu_\perp$ fulfilling Eq. (11) and Eq. (12). From Eq. (7) it can be proved that the permittivity and permeability of the final medium in spherical coordinates are respectively $\mathrm{diag}(\varepsilon_r, \varepsilon_\theta, \varepsilon_\phi)$ and $\mathrm{diag}(\mu_r, \mu_\theta, \mu_\phi)$, where

$$\varepsilon_r = \frac{\varepsilon_z}{r(1+\cos\theta)^2} \qquad \varepsilon_\theta = \frac{\varepsilon_\rho}{r} \qquad \varepsilon_\phi = \frac{\varepsilon_\psi}{r}$$
$$\mu_r = \frac{\mu_z}{r(1+\cos\theta)^2} \qquad \mu_\theta = \frac{\mu_\perp}{r} \qquad \mu_\phi = \frac{\mu_\perp}{r} \tag{17}$$

From the first of the transformation equations in Eq.(13), we can deduce that

$$\frac{2}{1+\rho^2} = 1+\cos\theta \tag{18}$$

And thus, using Eq. (12) we obtain:

$$\varepsilon_r = \frac{\varepsilon_z}{r(1+\cos\theta)^2} = \frac{1}{\mu_\perp r} \tag{19}$$

Since $\mu_\perp$ must be constant and $\varepsilon_z$ must be independent of z (we should better say that $\varepsilon_z$ must be independent of r, because Eq. (19) is written in coordinates r, θ, φ) then $\varepsilon_z$ should be proportional to $(1+\cos\theta)^2$. If we choose $\mu_\perp=1$, then $\varepsilon_z=(1+\cos\theta)^2=(2/(1+\rho^2))^2$. Now we can freely select $\varepsilon_\rho$, $\varepsilon_\psi$ and $\mu_z$ and calculate from Eq. (17) the components of ε and μ, without affecting the expression of the transformed field given by Eq. (16). This means that we can freely select $\varepsilon_\theta$, $\varepsilon_\phi$ and $\mu_r$, in the transformed medium. For instance, we can choose them to obtain an isotropic, impedance-matched final medium of refractive index n=1/r, with parameters:

$$\varepsilon_r = \varepsilon_\theta = \varepsilon_\phi = \mu_r = \mu_\theta = \mu_\phi = \frac{1}{r} \tag{20}$$

This medium has already been considered in [11] with a different purpose: by addition of a shell with the parameters of ε and μ also proportional to 1/r but negative, it was shown to perform as the spherical symmetric equivalent of Pendry's Perfect lens.
Other remarkable solutions exist when we select the function $f(r) = r$ in Eq. (13). One immediately arrives at an alternative medium (having the same perfect focusing properties) with the same refractive index n=1/r, which is also isotropic, but non-magnetic:

$$\varepsilon_r = \varepsilon_\theta = \varepsilon_\phi = \frac{1}{r^2} \qquad \mu_r = \mu_\theta = \mu_\phi = 1 \tag{21}$$

In contrast to the selection $f(r) = \ln(r)$, in the case $f(r) = r$ the mapping between the initial and final spaces is bijective only when the initial space is restricted to the domain $z>0$. This is not a problem, since we are specifically interested in the field solutions in limited domains of the input and output spaces, as discussed next.

## 3   Waveguides

The $z$-polarized waves of the initial medium are not perturbed by the addition of two perfect conductor plates at $z=$constant planes. Let's say $z=\ln R_1$ and $z= \ln R_2$, (see Fig. 2) since the electric field component tangent to that plane is null. These conductors create a waveguide within which the fields inside are isolated from the outside so that waves can propagate confined inside it. Correspondingly, the r-polarized fields in the final medium given by Eq. (16) propagate confined within the spherical waveguide formed by two perfect conductor concentric spheres $r = R_1$ and $r = R_2$. Due to transformation Eq.(13), the interior of that planar waveguide is transformed into the interior of the spherical waveguide and the flat conductors into the spherical conductors.

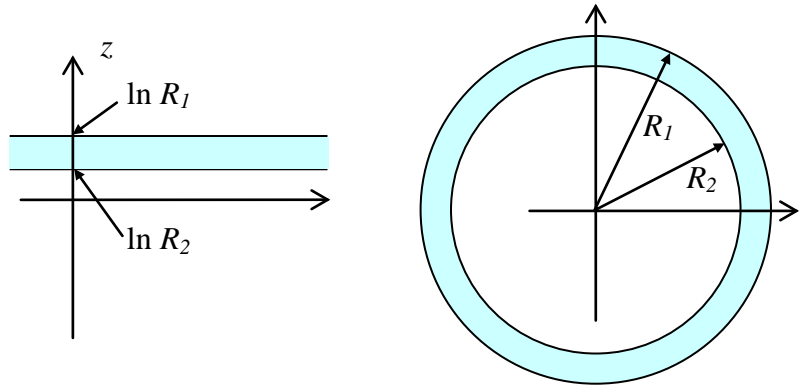

Fig. 2 Planar waveguide is transformed in a spherical one.

Let us consider the following specific time-harmonic solution of z-polarized wave in the initial medium described by Leonhardt in his Eq. (11) and (12) in [1]:

$$\mathbf{E} = E_z \mathbf{z} = \frac{P_\nu(\zeta) - e^{i\nu\pi} P_\nu(-\zeta)}{4\sin(\nu\pi)} e^{-i\omega t} \mathbf{z} \qquad (22)$$

Here $P_\nu$ is the Legendre function of the first kind,

$$\nu = \frac{-1 + \sqrt{1+4k^2}}{2} \qquad (23)$$

and

$$\zeta = \frac{\rho^2 - 1}{\rho^2 + 1} \qquad (24)$$

Such wave in the initial medium (Eq. (22)) is radial symmetric, generated by the line source located at the origin ($\rho=0$) and propagating outwards at the infinity ($\rho\to\infty$), where is drained. The line source is transformed by the first equation of Eq. (13) in the radial half-line $\theta=0$, while the line drain is transformed to the opposite radial half-line $\theta=\pi$. Also from the first equation of Eq. (13) we obtain that $\zeta =-\cos\theta$, which implies that the time-harmonic transformed field inside the spherical waveguide can be written from Eq. (16) and (22) as:

$$\mathbf{E}' = \frac{1}{r} \frac{P_\nu(-\cos\theta) - e^{i\nu\pi} P_\nu(\cos\theta)}{4\sin(\nu\pi)} e^{-i\omega t} \mathbf{r} \qquad R_1 \leq r \leq R_2 \qquad (25)$$

As a consequence, this field shows also an asymptotic behavior in the neighborhoods of the transformed source ($\theta$=0) and drain ($\theta$=$\pi$) as the one shown by the field in the original medium (Eq. (22)). Since the final medium is spherical symmetric, the source and drain absolute positions are arbitrary and so the field Eq. (25) is also obviously a solution after a rotation centered at the origin.

*3.1  Homogeneous waveguide*

The electric field of Eq. (25) inside the spherical waveguide has a $1/r$ dependence; the same as the refractive index $n=1/r$ of the filling medium. However, the range of variation of $r$ is limited to $R_1 \leq r \leq R_2$, so they are the field and refractive index variations. This implies that the closer $R_1$ and $R_2$ are selected, the closer the medium is to being homogeneous. Therefore, if a spherical waveguide is manufactured with $R_1$ and $R_2$ close enough, and is filled with an isotropic homogeneous material of refractive index $n=2/(R_1+R_2)$, its behavior is expected to be equally close to the MFE. Such proximity should lead to super-resolution properties, provided the MFE has the infinite resolution claimed by Leonhardt in [1].

**4     Generalization to a class of refractive index distributions: Geodesic waveguides**

In section 3 we found an isotropic non-magnetic medium wherein Helmholtz scalar waves can be replicated showing perfect imaging properties. Not only is this newly discovered medium isotropic and non-magnetic but we have also found that the waves can be confined in a waveguide within which the refractive index can be made almost constant. A key point of such a result is that the transformation from cylindrical variables $\rho$, $\psi$ to spherical variables $\theta$, $\phi$ is conformal. This keeps $\mu_\theta = \mu_\phi$ in the transformation medium, which is a necessary condition for the final medium to be isotropic. The transformation between any $z =$ constant plane of the original medium in their corresponding $r$=constant spheres is conformal. Nevertheless, we would only need the transformation between planes and spheres to be conformal within the spherical waveguide, *i.e.*, in $R_1 \leq r \leq R_2$.

In this section we will analyze other rotational symmetric geodesic waveguides. For some of the common refractive index distributions in a plane (i.e. 2D) with rotational symmetry $\eta(\rho)$ (such as the Eaton, and Gutman lenses besides the Maxwell Fish Eye) it is possible to establish a conformal mapping between points of the plane and points of a rotational symmetric surface so the geodesic curves on this surface are mapped in ray trajectories of a medium with refractive index $\eta(\rho)$ [12]. Let's call that surface, the geodesic surface corresponding to $\eta(\rho)$. Sometimes the mapping can not be established for all points of the plane. By placing two mirrors at two parallel surfaces placed at both sides of the geodesic surface and separated a small thickness $\tau$, it is possible to design a light guide with rays following trajectories close to the geodesic curves of the geodesic surface (by parallel surfaces we mean wavefront surfaces generated from a original surface by propagating it in a medium with constant refractive index, using geometrical optics). When $\tau \rightarrow 0$ the ray trajectories in the light guide are mapped into ray trajectories of the refractive index distribution $\eta(\rho)$. Obviously, the light guide corresponding to a MFE distribution is a spherical shell.

With the aid of transformation optics we will next show that waves with the electric field polarized normal to the waveguide surface are transformed from waves in a planar waveguide having a z-polarized electric field. This means that the mapping between geodesic waveguides and cylindrical refractive index distributions, a mapping valid in the Geometrical Optics framework, can be extended to certain modes of Wave Optics.

*4.1  The transformation*

The original medium is described in cylindrical coordinates $\rho$, $\psi$, $z$ and the final medium in a rotationally symmetric coordinate system $s$, $\sigma$, $\phi$. Similarly to Eq. (13) the three functions defining the transformation are

$$\rho = \rho(\sigma) \qquad \psi = \phi \qquad z = s \qquad (26)$$

For the surface defined by *s=0*, the coordinate $\sigma$ of a point **P'** in this surface is defined as the length of a meridian cross section curve from the axis to the point **P'** (see Fig. 3). The azimuthal angle is preserved in the mapping, *i.e.*, $\psi=\phi$. All the points in a straight line normal to *s=0* at **P'**, such as **Q'**, have the same value of $\sigma$ as **P'**. The coordinate *s* of a general point **Q'** is the length along these straight lines from the point to the intersection of the straight line with the surface *s*=0. We will assume that the Jacobian of the transformation defined by Eq. (26) is not singular in a region of the space around the plane *z=0* containing $-\tau/2 \le z \le \tau/2$ (or in its corresponding region around the surface *s=0*).

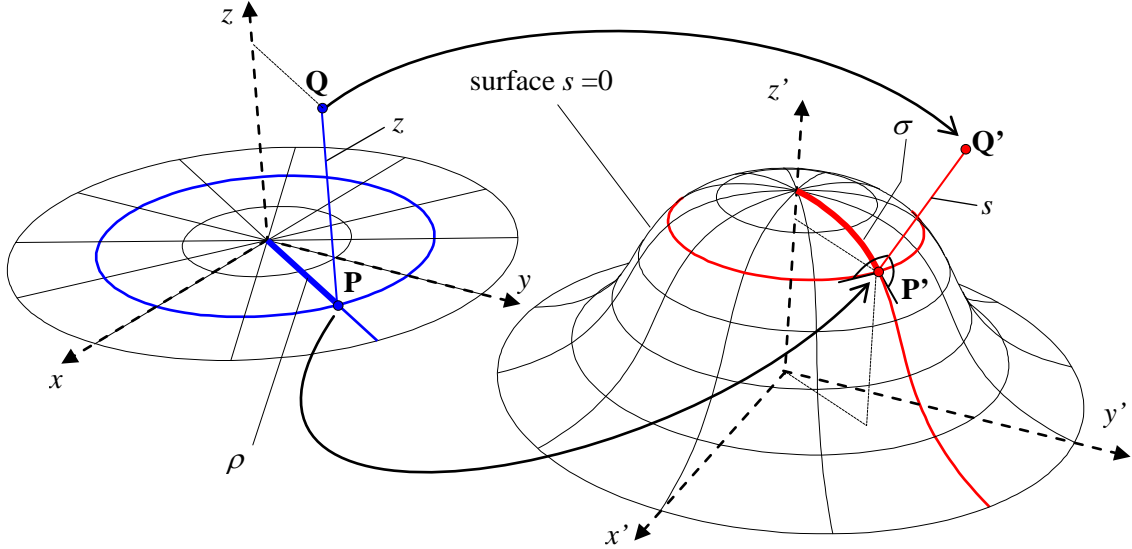

Fig. 3. Points of a plane are mapped onto a rotationally symmetric surface. The azimuthal angle is preserved in the mapping.

Note that with the above definition of coordinates, the scale factors $h_s$ and $h_\phi$ are $h_s=1$, $h_\phi=(x'^2+y'^2)^{1/2}$. The scale factor of $\sigma$ is $h_\sigma=1$, for the points of the surface *s=0*. Taking into account that the scale factors of cylindrical coordinates are $h_\rho=1$, $h_\psi=\rho$, $h_z=1$ the matrix M is

$$\mathrm{M} = \begin{pmatrix} \dfrac{h_s}{h_\rho}\dfrac{\partial s}{\partial \rho} & \dfrac{h_s}{h_\psi}\dfrac{\partial s}{\partial \psi} & \dfrac{h_s}{h_z}\dfrac{\partial s}{\partial z} \\ \dfrac{h_\sigma}{h_\rho}\dfrac{\partial \sigma}{\partial \rho} & \dfrac{h_\sigma}{h_\psi}\dfrac{\partial \sigma}{\partial \psi} & \dfrac{h_\sigma}{h_z}\dfrac{\partial \sigma}{\partial z} \\ \dfrac{h_\phi}{h_\rho}\dfrac{\partial \phi}{\partial \rho} & \dfrac{h_\phi}{h_\psi}\dfrac{\partial \phi}{\partial \psi} & \dfrac{h_\phi}{h_z}\dfrac{\partial \phi}{\partial z} \end{pmatrix} = \begin{pmatrix} 0 & 0 & 1 \\ h_\sigma \sigma_\rho & 0 & 0 \\ 0 & \dfrac{\sqrt{x'^2+y'^2}}{\rho} & 0 \end{pmatrix} \qquad (27)$$

If the surface *s=0* is a geodesic surface corresponding to the refractive index distribution $\eta(\rho)$, as explained before, then it fulfils (see for instance ref [12])

$$\sqrt{x'^2+y'^2} = \rho\,\eta(\rho) \qquad\qquad d\sigma = \eta(\rho)\,d\rho \qquad (28)$$

Then the matrix M is

$$\mathbf{M} = \begin{pmatrix} 0 & 0 & 1 \\ h_\sigma \eta(\rho) & 0 & 0 \\ 0 & \eta(\rho) & 0 \end{pmatrix} \qquad (29)$$

*4.2 Transformed fields*

According to Eq. (9), the electric field in the final medium is **E'**, as a function of the electric field in the initial medium **E** ($\mathbf{E}=E_\rho\boldsymbol{\rho}+E_\psi\boldsymbol{\psi}+E_z\mathbf{z}$). It is given by

$$\mathbf{E}' = E'_s\mathbf{s} + E'_\sigma\boldsymbol{\sigma} + E'_\phi\boldsymbol{\phi} = E_z\mathbf{s} + \frac{E_\rho}{h_\sigma\eta}\boldsymbol{\sigma} + \frac{E_\psi}{\eta}\boldsymbol{\phi} \qquad (30)$$

Since we are interested in the z-polarized waves ($E_\rho=0$, $E_\psi=0$), we obtain that the transformed field **E'** has only an s-component:

$$\mathbf{E}' = E_z\mathbf{s} \qquad (31)$$

*4.3 Permittivity and permeability tensors of the final medium*

Using Eq. (7) with $\varepsilon=\text{diag}(\varepsilon_\rho, \varepsilon_\psi, \varepsilon_z)$, $\mu=\text{diag}(\mu_\perp, \mu_\perp, \mu_z)$, we get that the permittivity and permeability of the final medium in $s$, $\sigma$, $\phi$ coordinates are respectively $\text{diag}(\varepsilon_s, \varepsilon_\sigma, \varepsilon_\phi)$ and $\text{diag}(\mu_s, \mu_\sigma, \mu_\phi)$, where

$$\varepsilon_s = \frac{\varepsilon_z}{h_\sigma\eta^2} \qquad \varepsilon_\sigma = h_\sigma\varepsilon_\rho \qquad \varepsilon_\phi = \frac{\varepsilon_\psi}{h_\sigma}$$
$$\mu_s = \frac{\mu_z}{h_\sigma\eta^2} \qquad \mu_\sigma = h_\sigma\mu_\perp \qquad \mu_\phi = \frac{\mu_\perp}{h_\sigma} \qquad (32)$$

Choosing again $\mu_\perp=1$ and $\varepsilon_z=\eta^2(\rho)$ (which fulfill the conditions of Eq. (11) of section 2.1), we can calculate $\varepsilon_s$, $\mu_\sigma$, $\mu_\phi$ from Eq. (32), and we can freely choose the remaining parameters $\varepsilon_\sigma$, $\varepsilon_\phi$, $\mu_s$, since $\varepsilon_\rho$, $\varepsilon_\psi$, $\mu_z$ do not intervene in the expression of the z-polarized field **E** in the original medium. This means that the transformed field **E'** (Eq. (31)) is insensitive to the selection of $\varepsilon_\sigma$, $\varepsilon_\phi$, $\mu_s$. For instance we can choose $\varepsilon_\sigma=\varepsilon_\phi=1/h_\sigma$, $\mu_s=1$ and get the following $\varepsilon$ and $\mu$ in the transformed medium:

$$\varepsilon_s = \varepsilon_\sigma = \varepsilon_\phi = \frac{1}{h_\sigma} \qquad \mu_s = 1 \qquad \mu_\sigma = h_\sigma \qquad \mu_\phi = \frac{1}{h_\sigma} \qquad (33)$$

*4.4 Homogeneous waveguide*

Let's design a waveguide with conductor surfaces $z=-\tau/2$ and $z=\tau/2$ (initial space) or $s=-\tau/2$ and $s=\tau/2$ (final space). In order for the fields to be well-defined, the Jacobian of the transformation Eq. (26) need to be non singular and single valued inside the waveguide, *i.e.*, between surfaces $s=-\tau/2$ and $s=\tau/2$. It can be easily checked that this is always achievable by a sufficiently small value of $\tau$.

Since $h_\sigma=1$ at the points of the surface $s=0$, the tensors $\varepsilon$ and $\mu$ at this surface are

$$\varepsilon_s = \varepsilon_\sigma = \varepsilon_\phi = \mu_s = \mu_\sigma = \mu_\phi = 1 \qquad (34)$$

This corresponds to a vacuum. Then, a sufficiently small $\tau$ makes the transformed medium nearly homogeneous, isotropic, and non-magnetic inside the waveguide. This means that when $\tau\rightarrow 0$, the z-polarized electric field mode in the initial medium can be reproduced in the waveguide filled with an isotropic, homogeneous, non-magnetic material, which seems much easier to manufacture.

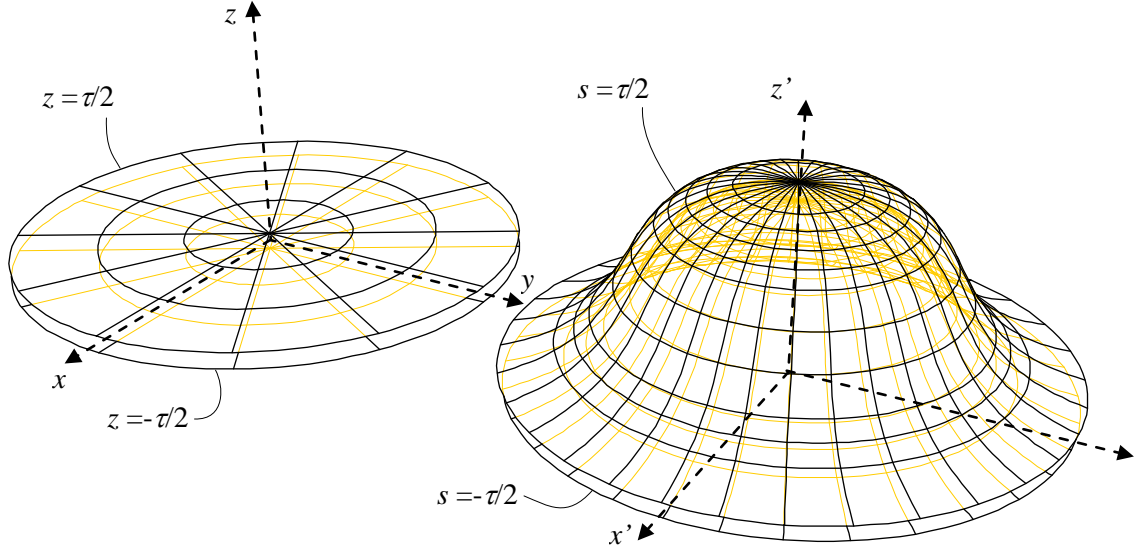

Fig. 4. The planar waveguide of the original space, which is filled with an inhomogeneous medium, is transformed into the curved waveguide filled with an isotropic, homogeneous, non-magnetic medium. The original *z*-polarized electric field is transformed into one polarized normal to the curved waveguide.

Besides the interest of the previous analysis of geodesic waveguides, the use of Transformation Optics seems to be a powerful tool to analyze waves in a curved waveguide giving much simpler results than the direct analysis [13].

## 5   Conclusions

With the help of Transformation optics we have been able to prove that:
1. A spherical waveguide filled with an isotropic non-magnetic material with radial refractive index $n=1/r$ has radial polarized modes with the same perfect focusing properties as the Maxwell Fish-Eye lens. Moreover, if the waveguide is thin enough, the medium within it can be approximated by a homogeneous one. These results pave the way to experimentally attain perfect imaging in the Maxwell Fish Eye lens
2. Geodesic waveguides designed with the tools of Geometrical Optics, from a 2D refractive index distribution, not only map rays in geodesic curves (as is well known from Geometrical Optics) but also replicate the *z*-polarized electric fields obtained in the cylindrical refractive index distribution of the waveguides.

**Acknowledgment**

The authors wish to thank Bill Parkyn for his help in editing and the Spanish Ministries MCINN (Engineering Metamaterials: CSD2008-00066, Deffio: TEC2008-03773, Sigmasoles: PSS-440000-2009-30) and MITYC (OSV: TSI-02303-2008-52), Madrid Regional Government (ABL: PIE/466/2009, F3: PIE/469/2009, SPIR: 50/2010) and UPM (R&D Groups) for the support given in the preparation of the present work.

**Appendix: Relationship between original and transformed electrical fields.**

Consider a transformation optics from a Cartesian space $x_1, x_2, x_3$ to a Cartesian space $x_1', x_2', x_3'$ composed of three transformations: from $x_1, x_2, x_3$ to $u_1, u_2, u_3$; from $u_1, u_2, u_3$ to $u_1', u_2', u_3'$; and from $u_1', u_2', u_3'$ to $x_1', x_2', x_3'$. The electrical fields in the original and transformed spaces are related by

$$\mathbf{E}_{x'} = \Lambda_{xx'}^{\mathrm{T}} \mathbf{E}_x \quad (35)$$

Here

$$\Lambda_{xx'} = \begin{pmatrix} \dfrac{\partial x_1}{\partial x'_1} & \dfrac{\partial x_1}{\partial x'_2} & \dfrac{\partial x_1}{\partial x'_3} \\ \dfrac{\partial x_2}{\partial x'_1} & \dfrac{\partial x_2}{\partial x'_2} & \dfrac{\partial x_2}{\partial x'_3} \\ \dfrac{\partial x_3}{\partial x'_1} & \dfrac{\partial x_3}{\partial x'_2} & \dfrac{\partial x_3}{\partial x'_3} \end{pmatrix} \quad (36)$$

The transformation is decomposed into three transformations

$$\Lambda_{xx'} = \Lambda_{xu} \Lambda_{uu'} \Lambda_{u'x'} \quad (37)$$

In order to calculate the fields in the spaces $u_1, u_2, u_3$ and $u_1', u_2', u_3'$ we will change variables from $x_1, x_2, x_3$ to $u_1, u_2, u_3$ and from $x_1', x_2', x_3'$ to $u_1', u_2', u_3'$. Denote by $\hat{\mathbf{E}}_u$ the normalized electrical fields in the space $u_1, u_2, u_3$ (and do similarly for $\hat{\mathbf{E}}_{u'}$).

$$\hat{\mathbf{E}}_u = \mathsf{H}_u \mathbf{E}_u \quad (38)$$

Here $\mathsf{H}_u$ is the matrix diag($h_1, h_2, h_3$) and $h_i$ are the scale factors. The fields in the original and final spaces are related by Eq. (35), and the normalized fields of any transformation are related by:

$$\hat{\mathbf{E}}_u = \Lambda_{xu}^{\mathrm{T}} \mathbf{E}_x \qquad \hat{\mathbf{E}}_{u'} = \Lambda_{uu'}^{\mathrm{T}} \hat{\mathbf{E}}_u \qquad \mathbf{E}_{x'} = \Lambda_{u'x'}^{\mathrm{T}} \hat{\mathbf{E}}_{u'} \quad (39)$$

Combining Eq. (37), (38), (39) gives

$$\mathbf{E}_u = \mathsf{H}_u^{-1} \Lambda_{u'u}^{\mathrm{T}} \mathsf{H}_{u'} \mathbf{E}_{u'} \quad (40)$$

This can also be written as

$$\mathbf{E}_u = \mathsf{M}^{\mathrm{T}} \mathbf{E}_{u'} \quad (41)$$